\documentclass[leqno,a4paper,12pt]{article}
\usepackage[centertags]{amsmath}
\usepackage{amsfonts}
\usepackage{amssymb}
\usepackage{multirow}
\usepackage{amsthm}
\usepackage{nicefrac}
\usepackage{pifont}
\usepackage[ansinew]{inputenc}
\usepackage{graphicx}

\theoremstyle{plain}

\newtheorem{rem}{Remark}[]

\begin{document}

\title{Surveillance of the Incidence of Noncommunicable Diseases (NCDs) with Prevalence Data: \\ Theory and Application to Diabetes in
Denmark}
\author{Ralph Brinks\\German Diabetes Center}
\maketitle

\begin{abstract}
Secular trends of the incidence of NCDs are especially important
as they indicate changes of the risk profile of a population. The
article describes a method for detecting secular trends in the
incidence from a series of prevalence data - without requiring
costly follow-up studies or running a register. After describing
the theory, the method is applied to the incidence of diabetes in
Denmark.
\end{abstract}

\section{Introduction}
Setting up health information systems to monitor the evolving
burden of noncommunicable diseases (NCDs) and their risk factors,
is one of the claims of the Moscow Declaration, which was approved
by the First Global Ministerial Conference on Healthy Lifestyles
and NCD control \cite{Mos11}. In that respect, secular trends of
the incidence of NCDs are especially important as they indicate
changes of the risk profile of the population under consideration.
Common ways to detect secular trends in the incidence are either
performing a series of follow-up studies or running a register.
Both approaches may be very costly and lead to a variety of
practical problems. In contrast, a series of prevalence studies
sometimes is much easier to accomplish. Based on illness-death
model (IDM), this article describes a method for detecting trends
in the incidence of NCDs without a series of follow-up studies and
without a register.

\bigskip

The next section introduces the IDM and derives the theoretical
background for the method. In the third section, the theory is
applied to data from the National Diabetes Register in Denmark.
The register observed an increasing diabetes incidence in
1995--2004. We show that the trend is detectable using the IDM and
a series of prevalence data. The last section contains a summary.

\section{Illness-Death Model}
\begin{figure}[ht]
  \centering
  \includegraphics[keepaspectratio,width=0.85\textwidth]{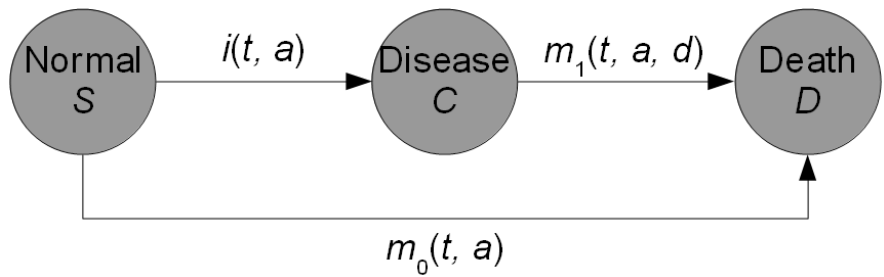}
\caption{Chronic disease model with three states and the
corresponding transition rates. People in the state \emph{Normal}
are healthy with respect to the disease under consideration. At
onset of the disease, they change to state \emph{Disease}.}
\label{fig:CompModel}
\end{figure}

In modelling chronic (i.e., irreversible) diseases, often the
three-state model (compartment model) in Figure
\ref{fig:CompModel} is used. The numbers of persons in the states
\emph{Normal} and \emph{Disease} are denoted by $S$ and $C$. The
transition intensities (synonymously: rates) between the states
are: the incidence rate $i$ and the mortality rates $m_0$ and
$m_1$ of the healthy or the diseased, respectively. These rates
generally depend on the calendar time $t$, the age $a$ and in the
case of the mortality $m_1$ also on the duration of the disease
$d$.

Although the inclusion of the disease duration $d$ is also
possible, hereinafter it is assumed that $m_1$ does not depend on
$d.$ This article analytically describes the relationship between
the prevalence of a chronic disease, the incidence and mortality
rates. This problem has existed at least since 1934 \cite{Mue34},
but so far has only been solved in special cases. The article
\cite{Hen10} presents a brief review and further references.

\bigskip

As in \cite{Bri12}, we look for the numbers $S(t, a)$ and $C(t,
a)$ of healthy and diseased persons in terms of differential
equations, which can be derived from the disease model in Figure
\ref{fig:CompModel}. For the healthy persons we get the following
initial value problem:
\begin{align}
(\partial_t + \partial_a) \, S & = - \left [ m_0 +
i \right ] \, S \label{e:PDE_S_ta} \\
S(t - a, 0) & = S_0(t - a). \nonumber
\end{align}

Here $S_0(t - a) = S(t - a, 0)$ is the number of (healthy)
newborns\footnote{This paper only considers diseases acquired
after birth.} at calendar time $t-a.$ The notation $\partial_x$
denotes the partial derivative with respect to $x, ~x \in \{t,
a\}$. The solution $S(t, a)$ is
\begin{equation}\label{e:S}
S(t, a)= S_0(t - a) \, \exp \left ( - \int_0^a m_0(t-a+\tau, \tau)
+ i(t-a+\tau, \tau) \, \mathrm{d}\tau \right ),
\end{equation}
which may be checked easily.

\bigskip

The number $C$ of diseased persons are described similarly:
\begin{align}
(\partial_t + \partial_a) \, C & = -m_1 \, C + i \, S \label{e:PDE_C_ta} \\
C(t, 0) & = 0. \nonumber
\end{align}

The solution is
\begin{equation}\label{e:C}
C(t, a) = \int_0^a i( t -\delta, a - \delta) \, S(t - \delta, a -
\delta) \exp \left ( - \int_0^\delta m_1(t - \delta + \tau, a -
\delta + \tau) \mathrm{d} \tau \right ) \mathrm{d}\delta.
\end{equation}

Equation \eqref{e:C} allows the following interpretation: Starting
from $(t, a)$ at $\delta$ time units before, i.e., at $(t -
\delta, a - \delta)$, exactly $i(t - \delta, a - \delta) \, S (t -
\delta, a - \delta)$ persons newly enter state \emph{Disease}.
Until $(t, a)$ the proportion
$$\exp \left (- \int_0 ^ \delta m_1 (t - \delta
+ \tau, a - \delta  + \tau) \mathrm{d} \tau \right)$$ of those has
survived. Integration over all possible $0 \le \delta \le a$
yields the number of diseased persons at time $(t, a).$ After
applying the quotient rule to the age-specific prevalence
$$p(t, a) = \frac{C (t, a)}{S (t, a) + C (t, a)}$$ and using
\eqref{e:PDE_S_ta} and \eqref{e:PDE_C_ta} it follows

\begin{align}
(\partial_t + \partial_a) \, p & = \left ( 1-p \right ) \, \left (
i  - p \, \left (m_1 - m_0 \right ) \right ) \label{e:PDE}\\
p(t, 0) & = 0. \nonumber
\end{align}

\begin{rem}\label{rem:nachI}
For $t, a \ge 0$ with $p(t, a) \neq 1$ it holds
\begin{equation}\label{e:inc}
    i(t, a) = \frac{(\partial_t + \partial_a) \, p (t,
    a)}{1-p(t,a)} + p(t, a) \, \left (m_1(t, a)
- m_0(t, a) \right ).
\end{equation}
\end{rem}

Furthermore, the solution of \eqref{e:PDE} can be calculated
directly via \eqref{e:S} and \eqref{e:C}:
\begin{equation}
p(t, a) = \frac{ \int \limits_0^a i(t - \delta, a - \delta) \,
\exp \left ( -\int\limits_0^\delta \Psi(t - \delta + \tau, a -
\delta + \tau) \mathrm{d} \tau \right )
\mathrm{d}\delta}{1+\int\limits_0^a i(t-\delta, a - \delta) \,
\exp \left ( -\int\limits_0^\delta \Psi(t - \delta + \tau, a -
\delta + \tau) \mathrm{d}\tau \right ) \mathrm{d}\delta},
\label{e:p}
\end{equation}
with $\Psi := m_1 - m_0 - i.$

\bigskip

This follows from
\begin{align*}
C(t, a)
&= \int_0^a i( t -\delta, a - \delta) \, S_0(t-a) \\ & \qquad \cdot \exp
\left ( - \int_0^\delta m_1(t - \delta + \tau, a - \delta + \tau)
\mathrm{d} \tau - \int_0^{a-\delta}(m_0 + i)(t - a + \tau, \tau)
\mathrm{d} \tau
\right ) \mathrm{d}\delta \displaybreak[0]\\
&= \int_0^a i( t - \delta, a - \delta) \, S_0(t-a) \\ & \qquad \cdot \exp
\left ( - \int_0^\delta m_1(t - \delta + \tau, a - \delta + \tau)
\mathrm{d} \tau - \int_0^{a}(m_0 + i)(t - a + \tau, \tau)
\mathrm{d} \tau \right. \\ & \qquad \qquad \qquad + \left.  \int_{a - \delta}^{a} (m_0 + i)(t - a + \tau, \tau)
\mathrm{d} \tau
\right ) \mathrm{d}\delta \displaybreak[0]\\
&= S_0(t-a) \, \exp
\left ( - \int_0^a (m_0 + i)(t-a+\tau, \tau) \mathrm{d}\tau
\right ) \\
& \qquad \cdot \int \limits_0^a i(t - \delta, a - \delta) \, \exp
\left ( - \int \limits_0^\delta \Psi(t - \delta + \tau, a - \delta
+ \tau) \mathrm{d} \tau \right ) \mathrm{d}\delta.
\end{align*}

The first part of the last expression equals $S(t, a)$ and
Equation \eqref{e:p} follows.

\bigskip

The usefulness of equation \eqref{e:p} is obvious: Given the
incidence $i(t, a)$ and mortalities $m_0(t, a), ~m_1(t, a)$, the
prevalence $p(t, a)$ can be calculated for all $t, a \ge 0$.

\bigskip

So we can state
\begin{rem}
The prevalence $p(t, a)$ is independent from the number $S_0$ of
newborns.
\end{rem}

\begin{rem}
For $t, a \ge 0$ it holds: $0 \le p(t, a) \le 1.$
\end{rem}

\begin{rem}\label{rem:increasing}
If for some $(t_1, a_1)$ the integral $$\Upsilon(t_1, a_1) :=
\int\limits_0^{a_1} i(t_1-\delta, a_1 - \delta) \, \exp \left (
-\int\limits_0^\delta \Psi(t_1 - \delta + \tau, a_1 - \delta +
\tau) \mathrm{d}\tau \right ) \mathrm{d}\delta$$ is lower than for
$(t_2, a_2): ~\Upsilon(t_1, a_1) < \Upsilon(t_2, a_2)$, then it
holds $p(t_1, a_1) < p(t_2, a_2).$ This follows from observing
that $x \mapsto \nicefrac{x}{1+x}, ~x\ge 0$ is strictly
increasing.
\end{rem}

\bigskip

At the end of the section we introduce the \emph{relative
mortality} $R(t, a).$ For $(t, a) \ge 0$ with $m_0(t, a) > 0,$
define
\begin{equation*}
        R(t, a) = \frac{m_1(t, a)}{m_0(t, a)}.
\end{equation*}

Now we have all the definitions and results for the next section.


\section{Diabetes in Denmark}
In the article \cite{Car08} the age-specific prevalence of
diabetes for men (and women) in Denmark in the period 1995-2007 is
presented in great detail. The results are based on a complete
survey of the Danish population ($n > 5$ million). Classifying a
person as diabetic is done by combining different health
registers, which yields a sensitivity of more than 85\% \cite[p.
2188]{Car08}. In this paper we confine ourselves to the male
population in Denmark. The age-specific incidence rate $i$ for
2004 is given for all age groups, but for the other years in the
period 1995-2007 just relatively to 2004, averaged across all age
groups; likewise, with the mortality $m_0$ of the non-diabetic
population. Mortality $m_1$ significantly depends on the disease
duration \cite[Fig. 4]{Car08}. To the apply the model of the
previous section model, the duration dependence has to be
suppressed. This is done by an initialization step: The relative
mortality $R^\star (a)$ is calculated such that the observed
incidence and the associated increase in prevalence from 1995 to
1996 are in agreement. Therefor, the Equation \eqref{e:PDE} is
solved for $m_1$. Then, $ R^\star$ is calculated by $R^\star =
\nicefrac{m_1}{m_0}.$ For the period 1996 - 2004, this relative
mortality is kept fixed and Equation \eqref{e:PDE} is solved for
$i$ as in Remark \ref{rem:nachI} with $m_1 = R^\star \, m_0.$ By
doing so, the relative mortality $R^\star$ for the period
1996-2004 is assumed to be independent from calendar time. Thus,
we have a two-step approach:
\begin{enumerate}
    \item Initialization: Calculate $R^\star$ by fitting the observed incidence rate in 1995
    and the increase in age-specific prevalence from 1995 to 1996.
    \item Application: Derivation of the incidence rates in 1996-2004
    via Remark \ref{rem:nachI} mit $m_1(t, a) = R^\star(a) \, m_0(t, a).$
\end{enumerate}

\begin{rem}
After initialization, we just use the mortality $m_0$ of the
non-diabetic population and the prevalences $p(t, a),$ $t=1996,
\dots, 2004,$ for deriving $i(t, a),$ $t=1996, \dots, 2004.$
\end{rem}

\begin{figure}[ht]
  \centering
  \includegraphics[width=.8\textwidth,keepaspectratio]{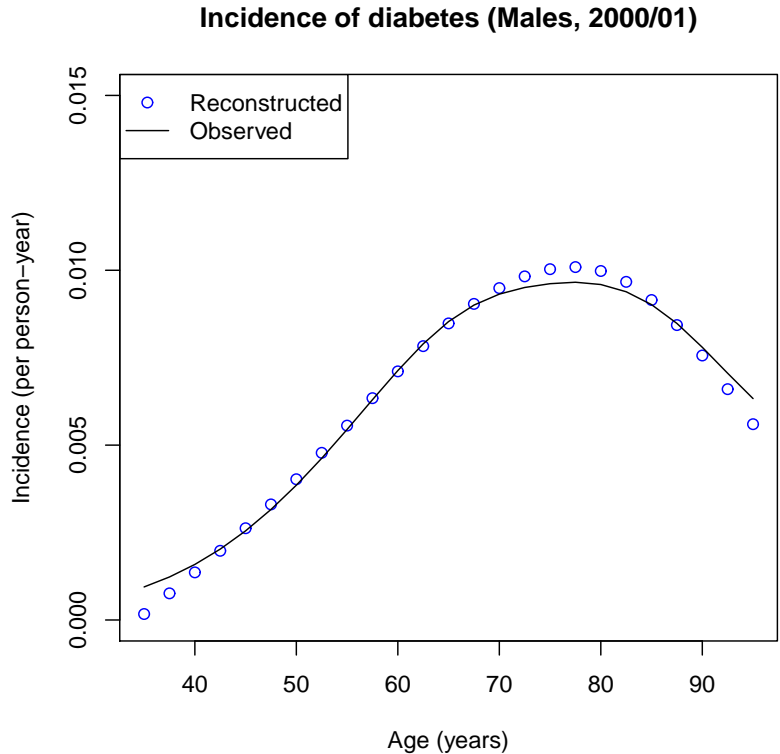}\\
  \caption{Age-spefic incidence rate in 2001: observed (solid
  line) and reconstructed by the model (circles).}\label{fig:Inc2001}
\end{figure}

Figure \ref{fig:Inc2001} shows the results of applying the model
to 2001 (circles). For comparison the observed incidence is shown
(solid line). Obviously, the data are in good agreement.

\bigskip

\begin{figure}[ht]
  \centering
  \includegraphics[width=.9\textwidth,keepaspectratio]{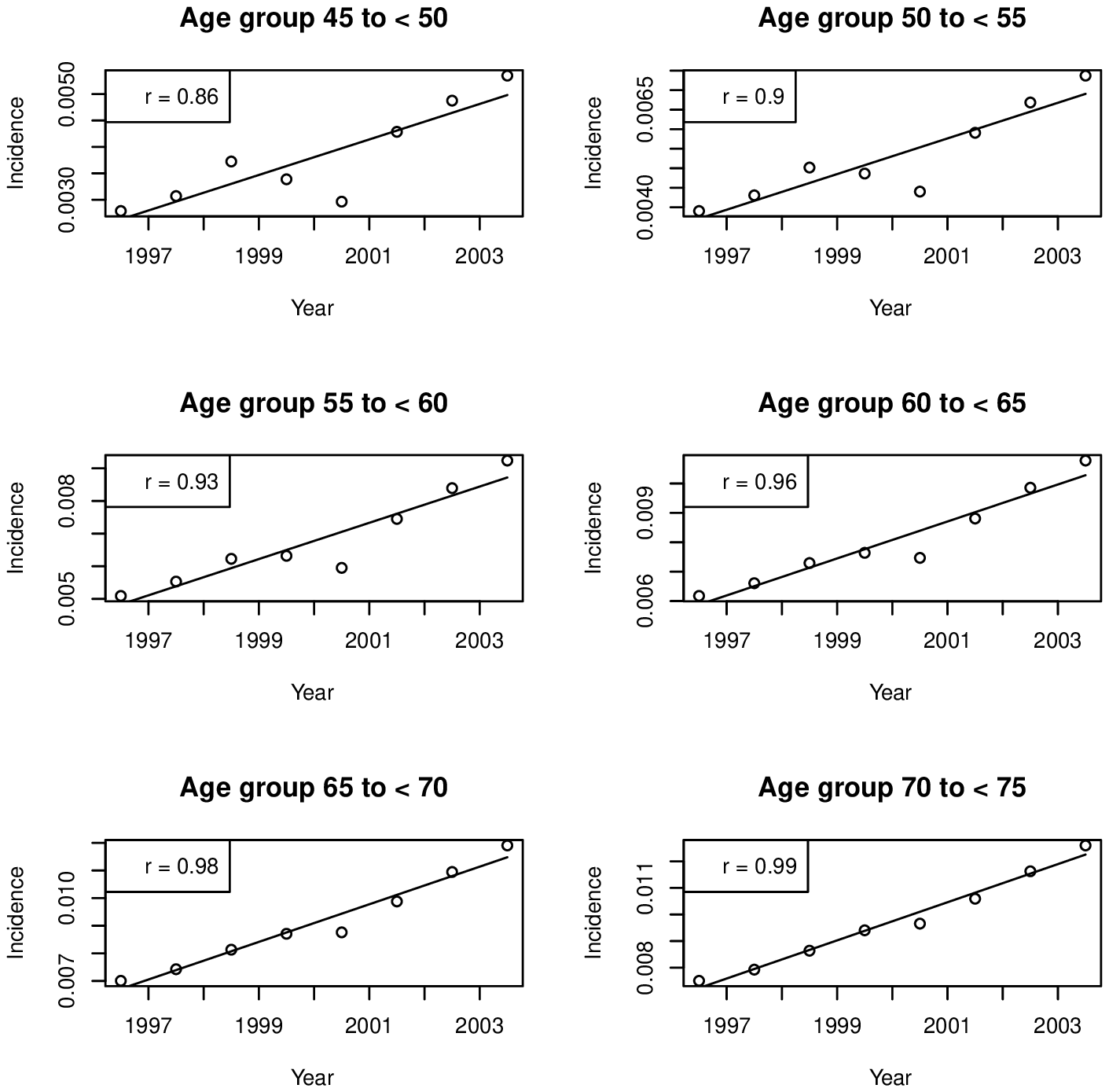}\\
  \caption{Incidence rates over calendar time for some age groups. }\label{fig:Trends}
\end{figure}

Now, the increase in the incidence can be examined. For some of
the age groups, the incidence rate over calendar time is shown in
Figure \ref{fig:Trends}. In addition, the regression lines and the
corresponding correlation coefficients $r$ are given. In all age
groups there is a significant upward trend. The higher the age,
the better the fit of the linear regression model. Table \ref{tab}
shows the numerical values for all age groups.

\begin{table}[ht]
  \centering
\begin{tabular}{c|c|c}
Age-    & Annual   & Correlation- \\
group   & increase (\%) & coefficient\\ \hline
40 -- 44 & 9.2 & 0.76 \\
45 -- 49 & 9.6 & 0.86 \\
50 -- 54 & 9.3 & 0.90 \\
55 -- 59 & 8.8 & 0.93 \\
60 -- 64 & 8.3 & 0.96 \\
65 -- 69 & 8.0 & 0.98 \\
70 -- 74 & 7.8 & 0.99 \\
75 -- 79 & 7.9 & 0.99 \\
80 -- 84 & 8.2 & 0.99 \\
\end{tabular}
\caption{Parameters of the secular trends in the incidence rates
in the age groups.}\label{tab}
\end{table}

The annual increase rates in the age groups (second column in
Table \ref{tab}) are all greater than the corresponding value
5.3\% reported in \cite[p. 2190]{Car08}. However, the reported
increase of 5.3\% refers to all persons (both sexes, all age
groups). Hence, a direct comparison with the values of Table
\ref{tab} is impossible.

\section{Summary}
In this work, a novel method for deriving trends in incidence from
a sequence of prevalence studies is presented. With a view to the
tremendous effort required by collecting incidence data, the novel
method provides a simpler alternative. A typical application is
the conversion of a sequence of telephone surveys for the
collection of age-specific prevalence of a chronic disease into a
sequence of incidence data.

As a first application, the method was used with data from the
Danish National Diabetes Register. The directly observed secular
trend in the incidence is visible by the new method as well.

\bigskip

In the application to the Danish Diabetes Register, the relative
mortality $R^\star$ in 1995 has been extrapolated for the period
1996-2004. While this may be possible for a period of eight years
is a word of caution is in order:
\begin{rem}
The calendar time trend in mortality $m_0$ of the non-diabetic
population is usually much better known than the trend in
mortality $m_1$ of the patients. The reason is that $m_0$ is
surveyed on a demographic scale, while $m_1$ is investigated
sporadically in epidemiological studies only. In epidemiology, one
might try to link the time trend in $m_1$ to the time trend in
$m_0$. The idea might be to measure a relative mortality $R$ at
some time $t' < t$ and extrapolate from $t'$ to $t$ and set $m_1
(t, a) = R (t', a) \, m_0 (t, a).$ Indeed, time dependence of
$m_1$ is enforced, but this approach may lead to a possibly
unexpected increase in the prevalence. If the incidence $i$ is
independent of $t$ and the family of functions $ t \mapsto m_{0,
a} (t): = m_0(t, a) $ is decreasing for all $a \ge 0$, then, by
Remark \ref{rem:increasing}, the function $t \mapsto p_{a}(t) : =
p (t, a)$ is monotonically increasing for all $ a \ge 0.$ This
means: although the incidence $i$ is remains unchanged by the
calendar time, by increasing the life expectancy of the healthy
(decreasing $m_{0, a}$), the prevalence increases. Extrapolating
the relative mortality $R$ from $t'$ to $t$ therefore must be
viewed critically.
\end{rem}

Beside the presentation of the theoretical background, this work
is little more than a feasibility study. There are two sources of
limitations:
\begin{enumerate}
    \item Data: Due to the incomplete detection of diabetes cases
    and the shortened report of incidence trends (pooled for all
    persons), a direct comparison between the observed and derived
    trends in the incidence is impossible.
    \item Model: Although it is evident that $m_1$ depends on the
    duration $d$ of diabetes, this dependency is neglected. In
    additon, the relative mortality $R^\star$ has been calculated
    for 1995, but has been extrapolated to 1996-2004.
\end{enumerate}
Both inaccuracies interact, which makes a rigorous evaluation
difficult. Thus, a systematic evaluation of the method based on a
comprehensive simulation study is necessary.


{}

\bigskip

\emph{Contact:} \\
Ralph Brinks \\
German Diabetes Center \\
Auf'm Hennekamp 65 \\
D- 40225 Duesseldorf\\
\verb"ralph.brinks@ddz.uni-duesseldorf.de"
\end{document}